\documentclass{INTERSPEECH2023}

\usepackage{amsmath,graphicx}
\usepackage{multicol,multirow}
\usepackage{url,cite}
\usepackage{hyperref}
\usepackage{xcolor}
\usepackage{tabularx,lipsum}
\usepackage{makecell}

\newcommand{\newpara}[1]{\vspace{8pt}\noindent\textbf{#1}}

\interspeechcameraready

\title{Curriculum learning for self-supervised speaker verification}
\name{
  \begin{tabular}{c}
  Hee-Soo Heo$^1$, Jee-weon Jung$^{1}$, Jingu Kang$^{2}$, Youngki Kwon$^1$,\\You Jin Kim$^1$, Bong-Jin Lee$^1$, Joon Son Chung$^3$
  \end{tabular}}
\address{
  $^1$NAVER Cloud Corporation, South Korea\\
  $^2$Bobidi, South Korea\\
  $^3$Korea Advanced Institute of Science and Technology, South Korea}
\email{heesoo.heo@navercorp.com}

\begin{document}

\maketitle
 
\begin{abstract}
The goal of this paper is to train effective {\em self-supervised} speaker representations without identity labels.
We propose two curriculum learning strategies within a self-supervised learning framework. 
The first strategy aims to gradually increase the number of speakers in the training phase by enlarging the used portion of the train dataset.
The second strategy applies various data augmentations to more utterances within a mini-batch as the training proceeds. 
A range of experiments conducted using the DINO self-supervised framework on the VoxCeleb1 evaluation protocol demonstrates the effectiveness of our proposed curriculum learning strategies. 
We report a competitive equal error rate of 4.47\% with a single-phase training, and we also demonstrate that the performance further improves to 1.84\% by fine-tuning on a small labelled dataset.
\end{abstract}

\section{Introduction}
Self-supervised learning (SSL) allows a model to map input data to a representative latent space without requiring human-annotated ground truth labels. 
Depending on the downstream task, models trained using self-supervision can serve as a pre-trained model or be used directly without further fine-tuning process~\cite{chen2021wavlm, wang2021unispeech, cai2021dku, cho2021jhu}. 
In both cases, its effectiveness is receiving attention, and various frameworks are being studied~\cite{grill2020bootstrap, baevski2022data2vec, he2021masked, chen2020improved}.

The speaker verification literature has also adopted SSL and several works have been proposed~\cite{stafylakis2019self, cho2020learning, ravanelli2019learning ,wu2021adversarial, cho2021jhu, vaessen2021fine, huh2020augmentation, xia2021self}.  
Specifically, few studies employed a two-phase self-supervised learning strategy, namely iterative clustering~\cite{tao2021self, cai2021dku}.
The first phase includes training the models via a SSL framework. 
The second phase repeats two steps until the performance converges to the intended level. 
First, pseudo labels are generated using the trained model. 
Second, the model is trained once again in a supervised classification manner, leveraging generated pseudo labels. 

Throughout this study, we focus on improving the self-supervised learning technique itself which involves randomly initialised models.
This line of research benefits the speaker verification literature in several aspects. 
First, it adopts single-phase training, saving a significant amount of time and does not require the estimation of the number of clusters.
Second, it coincides with developing other domains (e.g., image, natural language processing) where more advanced single-phase self-supervised frameworks are being studied. 
Third, if required, our model can serve as the initial model used for generating pseudo labels in iterative clustering, as our work corresponds to the first phase of it. 
Hence our improvements can complement the iterative clustering methods. 

Curriculum learning, which gradually trains a model in a meaningful order, is widely adopted in diverse supervised learning tasks~\cite{soviany2021curriculum, bengio2009curriculum, marchi2018generalised}.
Even when the same dataset is used, training with an adequately configured curriculum can boost the performance of trained models. 
However, curriculum learning has not been investigated in conjunction with SSL, leaving the potential open.
We thus focus on curriculum learning~\cite{bengio2009curriculum}.
Two curriculum learning strategies for SSL that make the training more challenging are designed: (i) gradually increasing the amount of training data and (ii) gradually augmenting noise and reverberation to an increased proportion within each mini-batch.
An underlying assumption is that SSL speaker verification will also benefit when the training becomes gradually difficult as it has been the case for a few preceding studies in supervised learning~\cite{chung2020defence}.

We conduct experiments on speaker verification with the ECAPA-TDNN model~\cite{desplanques2020ecapa} under the DINO~\cite{caron2021emerging} SSL framework.
Results demonstrate that SSL can also benefit from curriculum strategies. 
Both proposed curriculum learning techniques were effective, where we observed up to 33\% improvement compared to a baseline.
In addition, using models trained with SSL, we further explore a semi-supervised scenario where we fine-tune the model with a smaller set of data with labels.

The paper is organised as follows. 
Section~\ref{sec:curriculum} introduces conventional curriculum learning. 
The adopted SSL framework and model architecture, DINO, is addressed in Sections \ref{sec:dino}.
The proposed curriculum approach and techniques are addressed in Section \ref{sec:proposed}.
Experiments and result analysis are presented in Section~\ref{sec:exp}.
Section~\ref{sec:discussion} analyses and interprets the operation of the proposed technique in detail.

\section{Curriculum learning}
\label{sec:curriculum}

Curriculum learning, which defines the sequence of the training from the easy settings to the hard ones, allows efficient training of deep neural networks. 
Several configurations have been proved effective in the supervised learning field where performance improvements were observed with no additional overhead in terms of computations and resources~\cite{soviany2021curriculum, bengio2009curriculum}.
In \cite{marchi2018generalised}, the authors first trained the model under the text-dependent scenario, then extended to text-independent. 
Here, it has been shown that adjusting the training content through curriculum learning could teach the model to handle different textual content as well as make it robust in various acoustic environments. 
Other studies have also introduced methods of controlling training conditions.
Specifically, when adopting the additive angular margin loss function, 
increasing the margin as training proceeds has become a common technique~\cite{chung2020defence, huang2020curricularface}. 
As such, it has been demonstrated that the curriculum learning applied to the training condition stabilised the training and improved the quality of the model.

\section{Adopted SSL framework: DINO}
\label{sec:dino}

DINO~\cite{caron2021emerging} is a self-distillation framework based on the mean teacher~\cite{tarvainen2017mean} method, which was originally proposed for the computer vision domain. 
This framework employs a ``local-to-global'' distillation to guide the training of the student network. 
Precisely, various types of cropped and augmented views are constructed from an input, divided into local and global views depending on the resolution or the amount of information contained. 
Those with higher resolution or more information are the global views.
This framework aims to minimise the difference between the output features when different views are digested by either the teacher or the student network. 
In particular, the student network digests all views, then outputs local and global features, whereas the teacher network only digests global views, extracting global features.
Based on these two types of features, a loss $\mathcal{L}_{D}$ that penalises the difference between them is defined and used to train the student network.  
The loss can be described as:
\begin{equation}
\mathcal{L}_{D} = \Sigma_{\boldsymbol{x} \in \{ \boldsymbol{x}_{1}^{g}, \boldsymbol{x}_{2}^{g} \}} \Sigma_{\substack{\hat{\boldsymbol{x}} \in V  \\ \hat{\boldsymbol{x}} \neq \boldsymbol{x}} } H(P_{t}(\boldsymbol{x}), P_{s}(\hat{\boldsymbol{x}})),
\end{equation}
\noindent where $H(a,b)=-a \log b$, $\boldsymbol{x}_{i}^{g}$ indicates the global view, $V$ is the set of views from an input and $P_t(\cdot)$ and $P_s(\cdot)$ are the output distribution of teacher and student network, respectively. 
Different to the student which is trained with gradients, the teacher network's weight parameters are derived using an exponential moving average of the student. 
In addition, sharpening and centring techniques are applied to the teacher output.
The purpose is to avoid model collapse which can easily occur in frameworks that only utilise positive pairs, including DINO.

\begin{figure}[t!]
  \centering
  \includegraphics[width=0.85\linewidth]{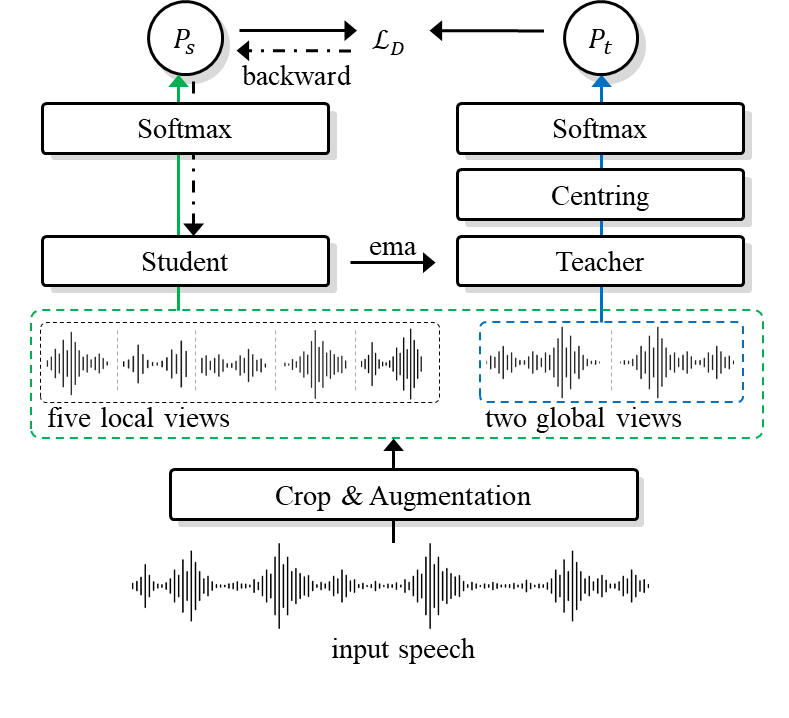}
  \caption{
    Adapted DINO framework for speaker verification.
    Utterances with different durations and augmentations are fed instead of different-sized images. Local and global views are fed into the student network (green line), while the teacher inputs only global views (blue line).
  } 
  \label{fig:DINO}
\end{figure}

We adapt DINO for speaker verification with a few modifications.
Figure~\ref{fig:DINO} illustrates the overall process of the DINO framework adapted for speaker verification. 
First, we change the global and local views to long and short crops of the same utterance with different augmentations. 
In particular, we construct two global views and five local views, resulting in seven views from each input. 
For augmentation, we use reverberation and noise from simulated RIRs and MUSAN datasets~\cite{Ko2017ARecognition,Snyder2015MUSAN:Corpus}. 
Augmentation configurations follow that of \cite{heo2020clova}.
Then, we control the sharpness of the student and teacher output distributions by using temperatures of $0.1$ and $0.04$, respectively, where the sharpness is controlled by dividing the output with temperature values before applying the softmax function. 
The aforementioned adaptation enables the speaker verification model training with the DINO framework. 
We followed the original paper for the other experimental configurations. 

\section{Proposed Approach}
\label{sec:proposed}
  
Existing curriculum learning strategies leverage the ground truth label. 
However, these strategies are not applicable for SSL because labels do not exist. 
Hence, we design two curriculum strategies that can be adopted in SSL without the ground truth label by increasing: (i) the size of the train set and (ii) the proportion of utterances within each mini-batch where data augmentation is applied. 
Both strategies tend to make the training progressively challenging.

\subsection{Data curriculum}
Finding a speaker discriminant latent space becomes more difficult as the number of speakers to represent increases. 
We wanted to gradually make the training more challenging by enlarging the number of speakers in the training dataset.
However, because labels do not exist in SSL, we could not manually control the number of speakers. 
As an alternative, we assume that the number of speakers in a dataset will increase proportionally when the size of the dataset enlarges and therefore we control the number of speakers by limiting the size of the training dataset.
Although this assumption cannot be guaranteed, a dataset that contains one million randomly collected utterances is likely to have a greater number of speakers compared to a dataset that contains one thousand utterances. 
In addition, in Section~\ref{sec:discussion}, we show another alternative where we adopt k-means clustering algorithm to select a subset of speakers' utterances.

Three following curriculum courses are designed empirically and illustrated in Figure \ref{fig:CL_DL}-(a) where it depicts the detailed ratios of data used for each epoch. 
For example, in the CL\_D2 strategy, we use the following dataset for training: 
During the first 16'th epochs, half of the dataset is used. 
From 17'th to 32'nd epochs, 75\% of the dataset is used. 
After the 32'nd epoch, entire dataset is used.
These curriculum courses are designed considering that the learning rate is reset every $16$ epochs in stochastic gradient descent with warm restarts (SGDR)~\cite{loshchilov2016sgdr} learning rate scheduler. 

\subsection{Data augmentation curriculum}
\begin{figure}[t!]
  \centering
  \includegraphics[width=0.85\linewidth]{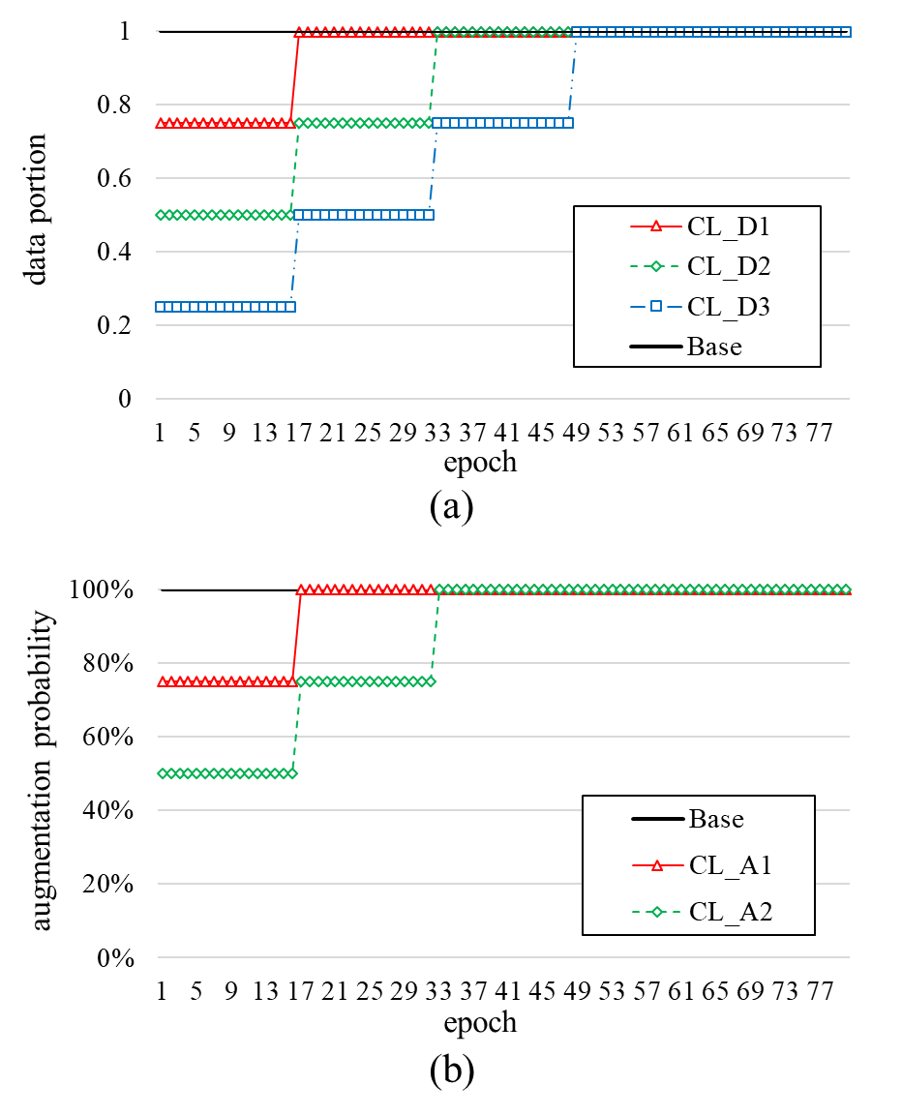}
  \caption{Curriculum courses.  Note that CL\_D* stands for \textbf{C}urriculum \textbf{L}earning on \textbf{D}ataset, which controls the amount of dataset used.  In a similar manner, in \textbf{C}urriculum \textbf{L}earning on \textbf{A}ugmentation (CL\_A*), we control the ratio of augmented data. (a): three courses for exploiting different data portions ({\em CL\_D1}, {\em CL\_D2} and {\em CL\_D3}) (b): two courses for augmenting a mini-batch ({\em CL\_A1} and {\em CL\_A2}).} 
  \label{fig:CL_DL}
\end{figure}

We further propose a curriculum strategy which adjusts the frequency of data augmentation to control how difficult the training would be.
We design two curriculum courses of augmentation by gradually increasing the proportion of augmented utterances within each mini-batch and illustrate specific curriculum courses in Figure \ref{fig:CL_DL}-(b). 
Note that the baseline augments all utterances within a mini-batch from the beginning of the training phase.

\section{Experiments}
\label{sec:exp}
We first present two sets of experiments: (i) demonstration of the effectiveness of proposed two curriculum strategies and (ii) fine-tuning the SSL-trained model with a small amount of labelled data (i.e., semi-supervised scenario).
Then, we compare our developed model's performance with the recent literature.

\subsection{Dataset}
\label{ssec:dataset}
Our experiments utilise the VoxCeleb1 and 2 datasets for training and evaluating the models~\cite{Nagrani19, Nagrani17, chung2018voxceleb2}. 
We use the development partition of the VoxCeleb2 dataset, which includes over a million utterances from $5,994$ speakers, to train the model with self-supervision, where we assume that the labels do not exist. 
The widely adopted equal error rate (EER) is the primary metric on the VoxCeleb1-O benchmark protocol. 

For the experiment of fine-tuning phase, the development partition of the VoxCeleb1 dataset which includes $148,642$ utterances from $1,211$ speakers and CN-Celeb~\cite{fan2020cn, li2022cn} are used.
We fine-tune the pre-trained model and evaluate using the corresponding test set using each data.
CN-Celeb is a dataset consisting of the voices of Chinese celebrities.
Among the entire set, we utilise CN-Celeb1, which includes $800$ speakers for training and $200$ speakers for evaluation.
Since CN-Celeb only comprises Chinese, it is valuable for evaluating the scenario where SSL pre-training is done in another domain.

\subsection{ECAPA-TDNN model}
\label{ssec:ecapa}
The model is based on the ECAPA-TDNN architecture that operates on mel-frequency cepstral coefficients input~\cite{desplanques2020ecapa}. 
It comprises a 1-dimensional convolution block followed by three Res2Net-based residual blocks with gradually increasing dilation values where a squeeze-excitation module is applied after each block. 
Up to the last residual block output, the sequence length remains because both pooling layers and stride size bigger than one are not included.
Three residual block outputs are concatenated and fed to a convolution block. 
Then, a context and channel-dependent statistical pooling layer aggregates frame representations into a single utterance representation. 
Finally, an affine transform derives the speaker embedding.

\subsection{Configurations}
\label{ssec:config}
Table~\ref{tab:config} describes the hyperparameters we use to train the model with the DINO framework.
For curriculum learning, we control the difficulty of training following settings shown in Figure \ref{fig:CL_DL}.
Since the curriculum strategies of augmentation and the data partition are implemented independently, they can be applied separately or together.
When two different curricula are applied simultaneously, the amount of data and the frequency of augmentation decrease. 

In the fine-tuning phase, the initial model is either randomly initialised, pre-trained using the DINO framework, or pre-trained via the DINO framework with proposed curriculum learning strategies. 
The fine-tuning of the model is accomplished utilising an open-source trainer~\footnote{ 
\url{https://github.com/clovaai/voxceleb_trainer}}, with the following modifications. 
We use additive angular margin loss to optimise the model~\cite{deng2019arcface, heo2020clova, liu2019large}. 
Most of the settings are used as the same with self-supervised learning, but the total number of epochs is reduced to $50$, and the learning rate scheduler is changed to SGDR without restart. 
\begin{table}
\centering
  \caption{Hyperparameters of the DINO framework.} 
\begin{tabular}{lc}
\Xhline{1pt}
\textbf{Configuration}           & \textbf{Value}                                                                                        \\
\Xhline{1pt}
optimizer               & adam                                                                                         \\
initial learning rate   & 0.001                                                                                        \\
weight decay            & 5.00E-05                                                                                     \\
batch size              & 200                                                                                          \\
epoch                   & 80                                                                                           \\
\begin{tabular}[c]{@{}l@{}}learning rate scheduler\end{tabular} & \multicolumn{1}{c}{\begin{tabular}[c]{@{}c@{}}SGDR~\cite{loshchilov2016sgdr}\\(restart period=16, decay=0.8)\end{tabular}}  \\
\Xhline{1pt}
\end{tabular}

  \label{tab:config}
  \vspace{-5pt}
\end{table}

\subsection{Results and Analysis}
\label{ssec:results}

\newpara{Self-supervised learning.}
Table~\ref{tab:SSL} addresses the effect of the two proposed curriculum learning strategies with the DINO framework.
Both curriculum strategies, regardless of combinations, consistently outperform the baseline.
However, although both approaches are effective when applied alone, they were not synergetic when applied simultaneously. 
The best performance was observed when only data curriculum was applied, CL\_D3, where it brought $30$\% improvement over the baseline.
We interpret these results that applying two kinds of curriculum at the same time lowers the difficulty of training beyond our expectation.

\begin{table}
\centering
\caption{Results of self-supervised learning in EER (\%) on the VoxCeleb1 test set. `Base' indicates the results from the DINO framework without curriculum learning. }
\begin{tabular}{lcccc}
    \Xhline{1pt}
       & Base & CL\_D1 & CL\_D2 & CL\_D3  \\
    \Xhline{1pt}
Base   & 6.70 & 5.87   & 5.54   & \textbf{4.47}    \\
CL\_A1 & 6.35 & 6.08   & 5.10   & 4.69    \\
CL\_A2 & 6.64 & 5.99   & 5.39   & 4.85    \\
    \Xhline{1pt}
\end{tabular}

\label{tab:SSL}
\end{table}  
\newpara{Semi-supervised learning.}
Table~\ref{tab:SL} shows the results of fine-tuning the best performing SSL-trained model, CL\_D3, with two datasets. 
The first row reports the results from a randomly initialised model that is trained only using the small labelled dataset.
In both datasets, replacing the random initialised model to SSL-trained model improved the performance, where 14.65\% and 11.87\% improvement were observed for VoxCeleb1 and CN-Celeb1.
By adopting the initial model trained with SSL under proposed curriculum strategies, further improvement was made, where 7\% and 3\% additional improvement were achieved.
Hence, we conclude that training in SSL with the proposed curriculum strategies are also effective for semi-supervised scenarios as well.

\newpara{Comparison with recent literature.}
Table \ref{tab:literature_ssl} compares the proposed models with existing works that adopt various self-supervision frameworks. 
Note that all these results show the performances of initial training without the iterative clustering step.
First, we find that the mainstream of SSL in speaker verification is on its transition from contrastive-based~\cite{huh2020augmentation,xia2021self,mun2020unsupervised,tao2021self} to DINO~\cite{han2022self,cho2022non}, which only leverages positive pairs.
Performances differ in each study, but in general, DINO outperforms contrastive-based approaches by a large margin.
Among studies that adopt DINO, the performance of our baseline model (6.70\%) falls behind a bit. 
However, with the best-performing curriculum learning strategy (DINO+CL), EER is further reduced to $4.47$\%, which is competitive.
Furthermore, since adapting DINO for speaker verification is subject to hyperparameter tuning, we argue that our improvements with curriculum strategies will further improve the performance of \cite{han2022self,cho2022non}. 

\begin{table}[t!]
\centering
\caption{Results of semi-supervised learning in EER (\%) on the VoxCeleb1 and CN-Celeb1 test set. Each model is fine-tuned by using the corresponding development set. }
\begin{tabular}{lcc}
\Xhline{1pt}
 \textbf{Initial model} &  \textbf{VoxCeleb1} &  \textbf{CN-Celeb1}  \\
\Xhline{1pt}
No pretraining          & 2.32      & 12.46         \\
\hline
DINO                    & 1.98      & 10.98         \\
DINO + CL               & \textbf{1.84}      & \textbf{10.65}         \\
\Xhline{1pt}
\end{tabular}
\label{tab:SL}
\vspace{-10pt}
\end{table}
 \begin{figure}[b!]
   \centering
   \includegraphics[width=0.85\linewidth]{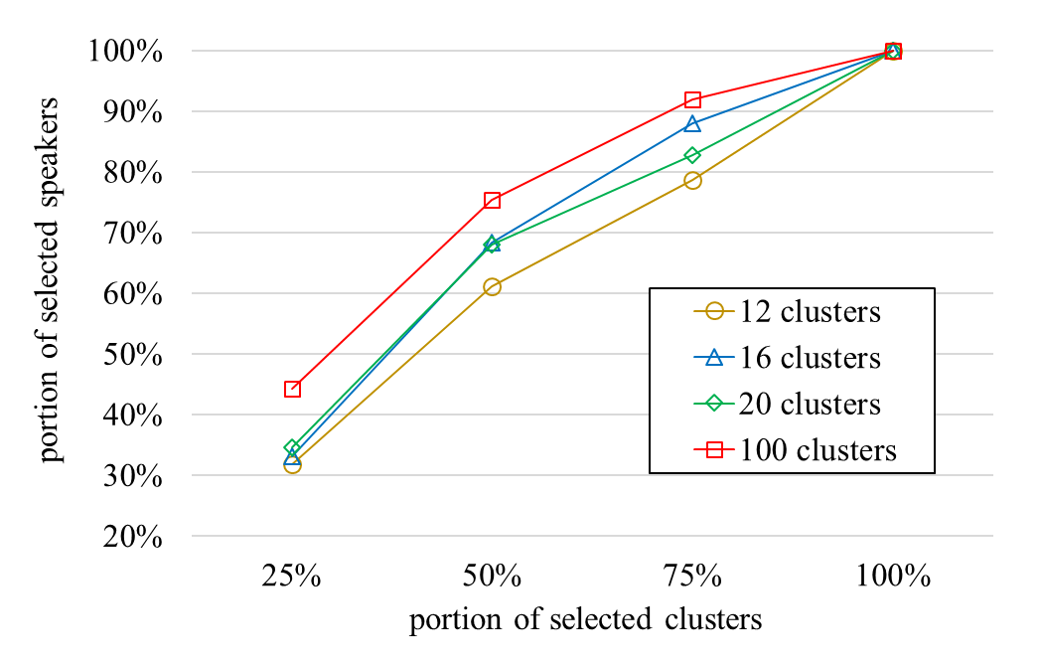}
   \caption{The proportion of speakers selected according to the number of randomly selected clusters.} 
   \label{fig:spkbycluster}
   \vspace{-10pt}
\end{figure}

\begin{table}[t]
 \caption{Comparison with self-supervised learning models. minDCF is calculated with $P_{target}$=0.05 and $C_{false \_alarm}=C_{miss}$=1.}
  \centering
  \label{tab:literature_ssl}
  \begin{tabular}{lccc}
  \Xhline{1pt}
    & \textbf{Framework} & \textbf{EER(\%)} & \textbf{minDCF}\\
  \Xhline{1pt} 
  Huh et al.~\cite{huh2020augmentation} & AP+AAT & 8.65 & 0.4540\\
  \multirow{2}{*}{Xia et al.~\cite{xia2021self}} & MOCO+Wav- & \multirow{2}{*}{8.23} & \multirow{2}{*}{0.5900}\\
  & Aug(ProtoNCE) & & \\
  Mun et al.~\cite{mun2020unsupervised} & CEL & 8.01 & N/R\\
  Tao et al.~\cite{tao2021self} & Contrastive & 7.36 & N/R\\
  Sang et al.~\cite{sang2021self} & SSReg & 6.99 & 0.4340\\
  Han et al.~\cite{han2022self} & DINO & 6.16 & N/R \\
  Cho et al.~\cite{cho2022non} & DINO & 4.83 & N/R \\
  \hline
  Ours & DINO & 6.70 & 0.4116 \\
  Ours & DINO+CL & 4.47 & 0.3057 \\
  \Xhline{1pt}
  \end{tabular}
  \vspace{-10pt}
\end{table}
\section{Discussion}
\label{sec:discussion}

\subsection{Analysis}
In SSL, it has been reported that rapidly increasing the representation power of the model is crucial. 
This is more important for SSL frameworks such as DINO, which relies entirely on positive pairs because representation collapse occurs more often.
We analyse that this may be the reason why curriculum learning was successful when applied to DINO SSL framework.
Our two curriculum learning strategies both initiates the training with easy samples. 
Therefore, the train loss can decrease and the quality of supervision can improve more rapidly.

\subsection{Ablation on CL\_D3}
\label{ssec:abl_cld3}
We conduct an ablation experiment to analyse which aspect brought the success of CL\_D3 in Table~\ref{tab:SSL}.
We hypothesised that gradually increasing inter-speaker variance is the key, and we design an experiment to validate this idea by removing the inter-speaker factor and only increasing intra-speaker variance over time. 
To do so, we use labels to ensure that although the same proportion of the training data is fed to train, the number of speakers are identical throughout the whole training phase\footnote{Labels are utilised for analysis purpose only.}. 
This experiment resulted in an EER of $6.41$\%, which is close to the EER of 6.70\% from DINO without curriculum.
We hence conclude that limiting inter-speaker variance at the initial phase of training is crucial to success.

\subsection{Additional method for controlling the number of speakers}
With the results of Section~\ref{ssec:abl_cld3}, a new concern can be made: the proposed data curriculum strategy may not hold if the number of speakers do not increase as the amount of collected data increase.
To account for such scenarios, we additionally demonstrate an alternative approach. 
A simple k-means clustering with given number of clusters is first adopted to group the train dataset.
Then, we select a proportion of clusters according to the curriculum strategy. 
Figure \ref{fig:spkbycluster} shows the experimental results. 
Regardless of the number of total clusters we set, we could successfully control and increase the number of speakers by choosing more clusters.
Based on this result, we conclude that even in the case where random selection of more data does not lead to the involvement of more speakers in the train dataset, we can apply a k-means algorithm alternatively and control the number of speakers.
Note that label information is not required for this process.

\section{Conclusion}
\label{sec:conclusion}
In this paper, we proposed two curriculum learning strategies for self-supervised speaker recognition.
The two strategies both demonstrated consistent improvements across a diverse range of experimental settings.
We also showed that the proposed methods are valid in a semi-supervised scenario.
In addition, in-depth analyses regarding the proposed curriculum strategies have been conducted.

\clearpage

\bibliographystyle{IEEEtran}
\bibliography{shortstrings, mybib}

\end{document}